# Combining Many-objective Radiomics and 3-dimensional Convolutional Neural Network through Evidential Reasoning to Predict Lymph Node Metastasis in Head and Neck Cancer

**Liyuan Chen[1,2], Zhiguo Zhou[1,2], David Sher[1], Qiongwen Zhang[1,2,3], Jennifer Shah[1], Nhat-Long Pham[1], Steve Jiang[1,2], and Jing Wang[1,2*]**

1. Department of Radiation Oncology, the University of Texas Southwestern Medical Center, Dallas, TX 75390 USA
2. Medical Artificial Intelligence and Automation (MAIA) Lab, the University of Texas Southwestern Medical Center, Dallas, TX 75390 USA
3. Department of Head and Neck Cancer, Cancer Center, West China Hospital, Sichuan University, Chengdu, China.

Email: Jing. Wang@UTSouthwestern.edu

**Abstract**

Lymph node metastasis (LNM) is a significant prognostic factor in patients with head and neck cancer, and the ability to predict it accurately is essential to optimizing treatment. Positron emission tomography (PET) and computed tomography (CT) imaging are routinely used to identify LNM. Although large or highly active lymph nodes (LNs) have a high probability of being positive, identifying small or less reactive LNs is challenging. The accuracy of LNM identification strongly depends on the physician's experience, so an automatic prediction model for LNM based on CT and PET images is warranted to assist LMN identification across care providers and facilities. Radiomics and deep learning are the two promising imaging-based strategies for node malignancy prediction. Radiomics models are built based on handcrafted features, while deep learning learns the features automatically. To build a more reliable model, we proposed a hybrid predictive model that takes advantages of both radiomics and deep learning based strategies. We designed a new many-objective radiomics (MO-radiomics) model and a 3-dimensional convolutional neural network (3D-CNN) that fully utilizes spatial contextual information, and we fused their outputs through an evidential reasoning (ER) approach. We evaluated the performance of the hybrid method for classifying normal, suspicious and involved

*Asterisk indicates corresponding author (e-mail: Jing.Wang@ UTSouthwestern.edu).
Liyuan Chen and Zhiguo Zhou contributed equally to this work.

LNs. The hybrid method achieves an accuracy (ACC) of 0.92 while XmasNet and Radiomics methods achieve 0.79 and 0.79, respectively. The hybrid method provides a more accurate way for predicting LNM using PET and CT.

# 1. INTRODUCTION

Lymph node metastasis (LNM) is a well-known prognostic factor for patients with head and neck cancer (HNC), which is the sixth most common malignancy worldwide [1]. LNM negatively influences overall survival and increases the potential of distant metastasis [2]. Radiation therapy is commonly used to control regional disease in the presence of nodal metastasis [3], where nodes of different malignant probability can be prescribed with different dose levels. Hence, accurately identifying LNM status is critical for therapeutic control and management of HNC. Cervical LNM status are routinely evaluated on positron emission tomography (PET) and computed tomography (CT), where PET provides functional activity of the LNM and CT provides high-resolution anatomical localization [4]. Although large or highly active lymph nodes (LNs), as identified by PET-CT, have a high probability of being positive, identifying small or less reactive LNs is challenging. The accuracy of LNM identification strongly depends on the physician's experience, so an automatic prediction model for LNM based on CT and PET images would help to assist LNM identification across care providers and facilities.

Imaging-based classification can be categorized into two major strategies: handcrafted feature-based and feature learning-based strategies. Among the handcrafted feature-based models, radiomics has shown great potentials for classification [5]. Through extracting and analyzing a large number of quantitative features, radiomics has been applied successfully to solve various prediction problems, such as tumor staging [6], treatment outcome prediction [7], and survival analysis [8]. Huang et al. [9] developed a radiomics model to predict LNM in colorectal cancer. This model extracted features from CT images and used multivariable logistic regression to build the predictive model. This model aims to a two-class prediction and only uses the classification accuracy as the objective function. To build a more reliable model, our group developed a multi-objective radiomics model [10] that considered both sensitivity and specificity simultaneously as the objective functions during model training. For feature learning-based models, deep learning is a powerful method that has been used to build predictive models for cancer diagnosis. Sung et al. [11] explored the use of deep learning methods, such as the convolutional neural network (CNN), deep belief network, and stacked de-noising auto-encoder to predict lung nodule malignancy. Zhu et al. designed a new CNN model to predict survival in lung cancer [12]. Yang et al. [13] built a model that combined the recurrent neural network and multinomial hierarchical regression decoder to predict breast cancer metastasis.

As both handcrafted feature-based and feature learning-based models have yielded promising results, one practical challenge is to determine which model is best suited to predicting LNM status. Features extracted by the feature learning-based model might be sensitive to global translation, rotation and scaling [14] while handcrafted features such as intensity features are not. Manually extracted features and automatically learned features could be complementary [15, 16], so combining them may yield more stable predictive results. Hence, a strategy that combines both handcrafted and learning models is a desired choice to predict LNM.

In this work, we propose a hybrid model that combines many-objective radiomics (MO-radiomics) and three-dimensional convolutional neural network (3D-CNN) through evidential reasoning (ER) to predict LNM in HNC. Because our multi-objective radiomics model [10] can only handle binary problems, we propose a new MO-radiomics model to predict the three classes of lymph nodes: normal, suspicious, and involved. Our proposed model considers procedure accuracy (PA) and user accuracy (UA) in confusion matrix, in addition to sensitivity and specificity, as objectives. We also designed a 3D-CNN consisting of convolution, rectified linear units (ReLU), max-pooling, and fully connected layers to automatically learn both local and global features for LNM prediction. The final output was obtained by fusing the MO-radiomics and 3D-CNN model outputs through the ER approach [17].

## 2. Materials and methods

### 2.1. Patient dataset

The study included PET and CT images for 31 patients with HNC who had enrolled in the INFIELD trial (https://clinicaltrials.gov/ct2/show/NCT03067610) between 2016 and 2017 at UT Southwestern Medical Center. Pretreatment PET and CT images were exported from digital Picture Archiving Communication System (PACS). Nodal status for all trial patients was reviewed by a radiation oncologist and a nuclear medicine radiologist. Figure 1 shows one example of CT and overlapped CT&PET images of normal, suspicious, and involved nodes. These nodes were contoured on contrast-enhanced CT guided by PET. We trained the predictive model on the lymph nodes of the first 21 patients, which included 53 involved nodes, 39 suspicious nodes, and 30 normal nodes. Then, we validated the predictive model on the remaining 10 independent patients with 13 involved nodes, 9 suspicious nodes, and 17 normal nodes. We used a total of 122 nodes for training and 39 nodes for testing.

### 2.2. Model overview

The workflow of the hybrid model is illustrated in Figure 2. First, patches of size 48×48×32, which include nodes and their surrounding voxels, were extracted as inputs for the proposed 3D-CNN model, while the nodes themselves were extracted as inputs for the MO-radiomics models. Then, the two model outputs were fused by ER to obtain the final output.

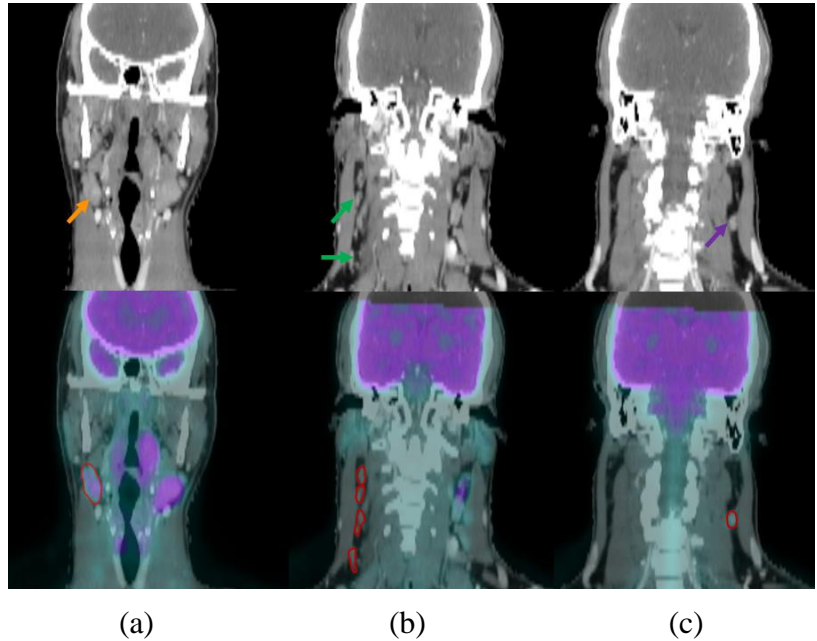

**Figure 1:** One example each of CT and overlapped CT & PET images of normal, suspicious, and involved nodes. Row 1: CT; Row 2: Overlapped CT & PET with contours of lymph nodes. (a)-(c) represent involved, suspicious and normal lymph nodes, respectively.

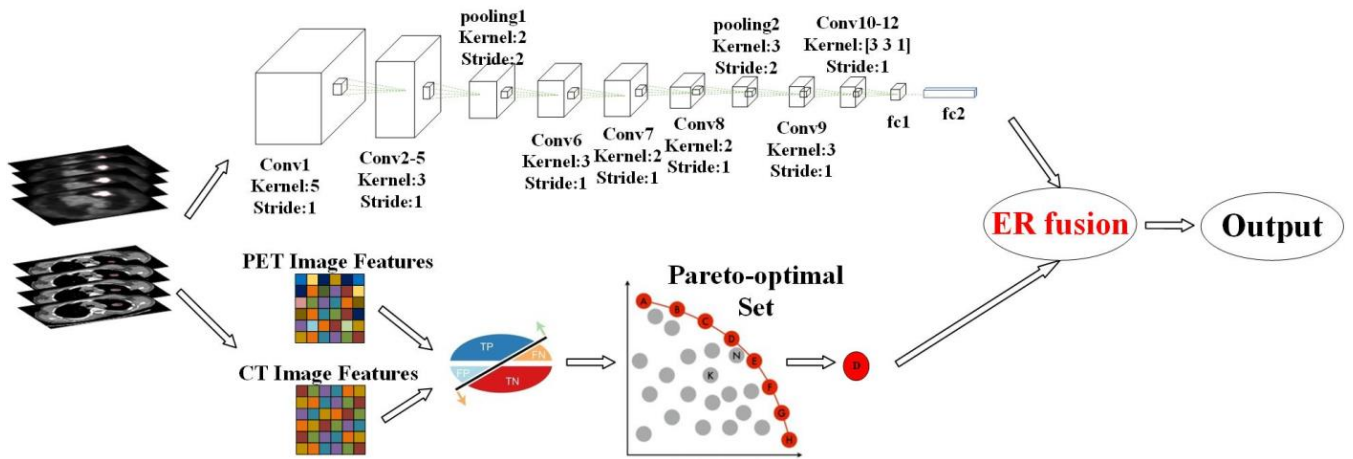

**Figure 2**: Workflow of the proposed hybrid model.

**2.3. MO-radiomics model**

In the MO-radiomics model, image features – including intensity, texture, and geometric features – are extracted from the contoured LNs (involved and suspicious) in PET and CT images. Additionally, at least one normal LN of similar size to the suspicious LNs was contoured to train the predictive model for each patient. Intensity features include minimum, maximum, mean, standard deviation, sum, median, skewness, kurtosis, and variance. Geometry features include volume, major diameter, minor diameter, eccentricity, elongation orientation, bounding box volume, and perimeter. Texture features based on 3D gray-level co-occurrence (GLCM) include energy, entropy, correlation, contrast, texture variance, sum-mean, inertia, cluster shade, cluster prominence, homogeneity, max-probability, and inverse variance. A total of 257 features were extracted for each PET and CT, respectively.

Then, we used the support vector machine (SVM) to build the predictive model with parameters denoted by $\alpha = \{\alpha_1, \cdots, \alpha_M\}$, where $M$ is the number of model parameters. All features, including PET and CT imaging features, are denoted by $\beta = \{\beta_1, \cdots, \beta_N\}$, where $N$ is the number of features. Procedure accuracy (PA) and user accuracy (UA) in confusion matrix were taken as objective functions because of the three classes of lymph nodes [18]. We maximized $f_{PA}^i$ and $f_{UA}^i$ simultaneously to obtain the Pareto-optimal set:

$$f = \max_{\alpha,\beta}(f_{PA}^i, f_{UA}^i, i = 1,2,3). \tag{1}$$

Equation (1) shows that six objective functions are considered in our model. The final solution of the selected features and model parameters can be selected from the Pareto-optimal set according to different clinical needs.

To solve the optimization problem defined in equation (1), we developed a many-objective optimization algorithm based on our previous multi-objective algorithm [10]. The proposed algorithm consists of two phases: (1) Pareto-optimal solution generation; and (2) best solution selection. The first phase is the same as in the multi-objective algorithm [10] which includes initialization, clonal operation, mutation operation, deleting operation, population update, and termination detection. In the second phase, the final solution is selected according to accuracy and AUC. Assume that the thresholds for accuracy are denoted by $T_{acc}$. The Pareto-optimal solution is denoted by $D = \{D_1, D_2, \cdots, D_P\}$. The corresponding accuracy and AUC for each individual $D_i, i = 1,,2,\cdots,P$ are denoted by $D_i^{acc}, D_i^{AUC}, i = 1,2,\cdots,P$, respectively. The procedure to select the best solution is as follows: *Step 1)* For each solution set $D_i, i = 1,2,\cdots,P$, if $D_i^{acc} > T_{acc}$, $D_i$ is selected. All selected candidates constitute the new candidate set denoted by $D_C = \{D_C^1, D_C^2, \cdots, D_C^Q\}$, where $Q$ is the number of selected individuals, i.e., feasible solutions. *Step 2)* Solution with highest AUC in $D_C$ is selected as the final solution $P^*$.

**2.4. 3D-CNN model**

The architecture of the proposed 3D-CNN model consists of 12 convolutional layers, 2 max-pooling layers, and 2 fully connected layers (Figure 2). Each convolutional layer is equipped with ReLU activation [19] and batch normalization. The construction order of the different layers in the architecture is shown in Table 1.

**Table 1**: 3D-CNN architecture.

| Layer | Kernel Size | Stride | Output Size | Feature Volumes |
|---|---|---|---|---|
| **Input** | - | | 48×48×32 | 1(or 2) |
| **C1** | 5×5×5 | [1 1 1] | 44×44×28 | 32 (or 64) |
| **C2** | 3×3×3 | [1 1 1] | 42×42×26 | 64 |
| **C3** | 3×3×3 | [1 1 1] | 40×40×24 | 64 |
| **C4** | 3×3×3 | [1 1 1] | 38×38×22 | 64 |
| **C5** | 3×3×3 | [1 1 1] | 36×36×20 | 64 |
| **MP1** | 2×2×2 | [2 2 2] | 18×18×10 | 64 |
| **C6** | 3×3×3 | [1 1 1] | 16×16×8 | 64 |
| **C7** | 2×2×2 | [1 1 1] | 15×15×5 | 64 |
| **C8** | 3×3×3 | [1 1 1] | 13×13×4 | 64 |
| **MP2** | 3×3×3 | [2 2 2] | 6×6×2 | 64 |
| **C9** | 3×3×3 | [1 1 1] | 6×6×2 | 64 |
| **C10** | 3×3×1 | [1 1 1] | 6×6×2 | 64 |
| **C11** | 3×3×1 | [1 1 1] | 6×6×2 | 64 |
| **C12** | 3×3×1 | [1 1 1] | 6×6×2 | 32 |
| **FC1** | | | 1×1×1 | 256 |
| **FC2** | | | 1×1×1 | 3 |

*C indicates Convolution layer + ReLU layer +Batch Normalization layer; MP indicates Max-pooling layer; and FC indicates Fully connected layer.

Since the max-pooling layer provides basic translation invariance to the internal representation, the convolutional and max-pooling layers are arranged alternately in the proposed architecture. In addition, since the max-pooling layer down-samples the feature maps, the convolutional layers in the architecture can capture both local and global features. For the last four convolutional layers, we first pad zero around each feature map from previous layers and then perform the convolution to preserve the output size of each feature map, which guarantees deep extraction and analysis of the 3D image features. The second fully connected layer finally generates three predicted probabilities as output of the 3D-CNN model. Because we aim to extract information from both PET and CT simultaneously, the input consists of two volumetric images, each of which serves as a channel of the final 4D data input.

The key steps to train the proposed 3D-CNN model are as follows:

*Normalization of the input:* Inputs with same scale can make converge faster during network training. The CT number range for our obtained CT images varies from -1000 to +3095. Hence, the specific normalization formula for CT image in this work is as follows:

$$I^{CT}_{input} = (I^{CT} + 1000)/4095, \tag{2}$$

which makes CT input into the range of [0, 1]. For PET, we first calculated the standardized uptake value (SUV) for each patient. Then the input of PET image is normalized as follows:

$$SUV_{input} = SUV/SUV_{maximum,training}, \tag{3}$$

which also makes the PET input into the range of [0, 1]. Here, $SUV_{maximum,training}$ indicates the maximum SUV value of the training dataset. We applied this normalization strategy for both training and testing images.

*Data augmentation and balance:* Imbalanced data might affect the CNN model's efficiency [20]. Our training dataset had 53 involved, 39 suspicious, and 30 normal nodes. Hence, we increased the number of samples for the suspicious and normal classes. We used the Synthetic Minority Over-sampling technique (SMOTE) to generate synthetic examples for these two minority classes. Synthetic examples were introduced along the line segments joining all of the k minority class nearest neighbors for each minority class sample until we had 53 nodes for each class for training. Data augmentation has been proven effective for network training [21]. We rotated the 3D nodes along $x, y, z$ three dimensions by 30°, 45°, 60°, 75° to generate more training samples.

*Initialization of the 3D-CNN weights:* Initializing the network weights will affect the convergence of the network training. We use Xavier initialization in our network to guarantee that the variance of the input and output for each layer is the same to avoid back-propagated gradients vanishing or exploding, so that activation functions can work normally.

*Loss function:* We use categorical cross entropy as the objective function that our network minimizes for LNM prediction. The categorical cross entropy formula for our network is as follows:

$$H(p,q) = -\sum_x q(x)\log(p(x)), \tag{4}$$

where $q(x)$ is the target and $p(x)$ represents the predicted probabilities.

**2.5. Evidential reasoning fusion**

After obtaining the outputs from the MO-radiomics and 3D-CNN models, the final output is generated using ER. Assume that $P^1 = \{p_1^1, p_2^1, p_3^1\}$ represents the output of MO-radiomics, and the 3D-CNN output is denoted by $P^2 = \{p_1^2, p_2^2, p_3^2\}$. They satisfy the following constraint:

$$\sum_{i=1}^{3} P_i^j = 1, \quad 0 \le P_i^j \le 1, j = 1,2. \tag{5}$$

Given the weight $\omega = \{\omega_1, \omega_2\}$, which satisfies $\omega_1 + \omega_2 = 1, 0 \leq \omega_j \leq 1$, the final output $P_i, i = 1,2,3$ is obtained through the following equations:

$$P_i = \frac{\mu \times \left[\prod_{j=1}^{N}\left(\omega_j P_i^j + 1 - \omega_j \sum_{i=1}^{M} P_i^j\right) - \prod_{j=1}^{N}\left(1 - \omega_j \sum_{i=1}^{M} P_i^j\right)\right]}{1 - \mu \times \left[\prod_{j=1}^{N}(1-\omega_j)\right]}, \tag{6}$$

$$\mu = \left[\sum_{i=1}^{M} \prod_{j=1}^{N}\left(\omega_j P_i^j + 1 - \omega_j \sum_{i=1}^{M} P_i^j\right) - (N-1) \prod_{j=1}^{N}\left(1 - \omega_j \sum_{i=1}^{M} P_i^j\right)\right]^{-1}, \tag{7}$$

where $M = 3$ and $N = 2$. Finally, the label $L$ is obtained by:

$$L = \max(P_i). \tag{8}$$

## 2.6. Comparison methods

We compared the proposed hybrid model with a convolutional neural network-based method, XmasNet, which was recently proposed by Liu [22] and has been shown to be effective for prostate cancer diagnosis on Multi-parametric MRI. We also compared our hybrid model with the conventional radiomics method proposed by Vallières et al. [23], which only considers accuracy as the objective function during the model training. Additionally, we evaluated the performance of the proposed MO-radiomics and 3D-CNN methods separately to illustrate the effectiveness of the ER fusion technique in the hybrid model. In addition to combining PET and CT as input, we also used PET and CT alone to build the predictive models for each method.

## 2.7. Evaluation criteria

Since our hybrid model has to handle three categories of nodules (normal, suspicious and involved), we used five criteria – confusion matrix, accuracy (ACC), Macro-Average, mean-one-versus-all (OVA)-AUC, and multiclass AUC [24] – to evaluate its performance. A confusion matrix is a commonly used specific table layout (Table 2) that visualizes the performance of a supervised learning algorithm. Each row of the matrix represents the instances in a predicted class, and each column represents the instances in an actual class. Accuracy is measured by the ratio of number of correctly labelled samples to total number of samples. Macro-Average is defined as the average of the correct classification rates. This measure has been used as a simple way to handle more appropriately unbalanced datasets [25]. Mean-OVA-AUC is defined as the average of the one-versus-all AUCs, which can be used as a measure of how well the classifier separates each class from all the other classes. Multi-class AUC, proposed by Hand et al. [24], is an extended definition of two-class AUC that averages pairwise comparisons to evaluate multi-class classification problems.

**Table 2**: An example of Confusion Matrix. (Here, $N_{ab}$ represents the instances of actual A class predicted in the B class.)

|  |  | Predicted |  |  |
|---|---|---|---|---|
|  |  | A | B | C |
|  | A | $N_{aa}$ | $N_{ab}$ | $N_{ac}$ |
| Actual | B | $N_{ba}$ | $N_{bb}$ | $N_{bc}$ |
|  | C | $N_{ca}$ | $N_{cb}$ | $N_{cc}$ |

The formulas for calculating ACC, Macro-average, Mean-OVA-AUC, and multi-class AUC values for measuring three-class prediction are listed in Table 3.

**Table 3**: Formulas of the four evaluation criteria.

|  | Formula |
|---|---|
| ACC | $(N_{aa} + N_{bb} + N_{cc}) / \sum_{i,j \in [a,b,c]} N_{ij}$ |
| Macro-average | $\left[ \frac{N_{aa}}{\sum_{j \in [a,b,c]} N_{aj}} + \frac{N_{bb}}{\sum_{j \in [a,b,c]} N_{bj}} + \frac{N_{cc}}{\sum_{j \in [a,b,c]} N_{cj}} \right] / 3$ |
| Mean-OVA-AUC | $[AUC_{a\ vs.all} + AUC_{b\ vs.all} + AUC_{c\ vs.all}]/3$ |
| Multi-class AUC | $[\hat{A}(a,b) + \hat{A}(a,c) + \hat{A}(b,c)]/3$ |

In Table 3, $\hat{A}(i,j) = \frac{[A(i,j)+A(j,i)]}{2}, i,j \in [a,b,c]$ with $A(i,j)$ is the probability that a randomly drawn member of class $j$ will have a lower estimated probability of belonging to class $i$ than a randomly drawn member of class $i$. Based on the definitions of these four evaluation criteria, higher values indicate better prediction results.

**2.8. Implementation Details**

For the many-objective training algorithm, the population number was set to 100, while the maximal generation number was set to 200. The mutation probability was set to 0.9 in the mutation operation. For the 3D-CNN model training, the Adam optimization algorithm was used with learning rate as 1e-5.

**3. Results**

We summarized the performance of the five methods in predicting lymph node metastasis using CT and PET modality images by listing the ACC, Macro-average, Mean-OVA-AUC, and multi-class-AUC values obtained by each method (Table 4) and showing bar plots of the values of the four evaluation criteria in Fig. 3. The proposed CNN and MO-radiomics models always show better prediction results than the popular XmasNet and conventional radiomics models, whether using only CT or PET images or a combination of PET and CT images. For example, the proposed CNN model using the combination of PET and CT images as input outperformed the XmasNet or Radiomics models in prediction accuracy by 0.13 (around 15%). The hybrid method, which integrates the outputs of the proposed CNN and MO-radiomics models, obtained even or higher evaluation criteria values compared to the CNN or MO-radiomics model alone. Although the MO-radiomics and 3D-CNN models had the classification accuracy values of 0.76 and 0.82 respectively using CT imaging, the hybrid model improved the accuracy to 0.87, indicating the effectiveness of the ER fusion strategy. The Macro-average value obtained by using CT imaging was improved by the hybrid model to 0.86 from 0.69 and 0.82 for MO-Radiomics and CNN, respectively. The mean-OVA-AUC and multi-class-AUC values obtained by the hybrid model outperformed the two single models, suggesting that results are more reliable after combination. Since the proposed CNN model have already achieved a high accuracy of 0.92 (only 3 nodes were misclassified), mean-OVA-AUC of 0.97 and multi-class AUC of 0.96, the ER fusion strategy did not improve the classification further.

**Table 4**: Values of four evaluation criteria obtained by different methods.

| | | CT | PET | PET&CT |
|---|---|---|---|---|
| ACC | XmasNet | 0.71 | 0.76 | 0.79 |
| | Radiomics | 0.74 | 0.77 | 0.79 |
| | Proposed CNN | 0.82 | 0.87 | 0.92 |
| | Mo-radiomics | 0.76 | 0.79 | 0.82 |
| | Hybrid | **0.87** | **0.87** | **0.92** |
| Macro-average | XmasNet | 0.67 | 0.78 | 0.78 |
| | Radiomics | 0.72 | 0.77 | 0.80 |
| | Proposed CNN | 0.82 | 0.85 | 0.91 |
| | Mo-radiomics | 0.69 | 0.80 | 0.82 |
| | Hybrid | **0.86** | **0.86** | **0.91** |
| Mean-OVA-AUC | XmasNet | 0.84 | 0.85 | 0.91 |
| | Radiomics | 0.84 | 0.92 | 0.90 |
| | Proposed CNN | 0.92 | 0.95 | 0.97 |
| | Mo-radiomics | 0.91 | 0.93 | 0.90 |
| | Hybrid | **0.93** | **0.95** | **0.97** |
| Multi-class-AUC | XmasNet | 0.83 | 0.85 | 0.89 |
| | Radiomics | 0.81 | 0.92 | 0.89 |
| | Proposed CNN | 0.92 | 0.95 | 0.96 |
| | Mo-radiomics | 0.89 | 0.93 | 0.89 |
| | Hybrid | **0.93** | **0.95** | **0.96** |

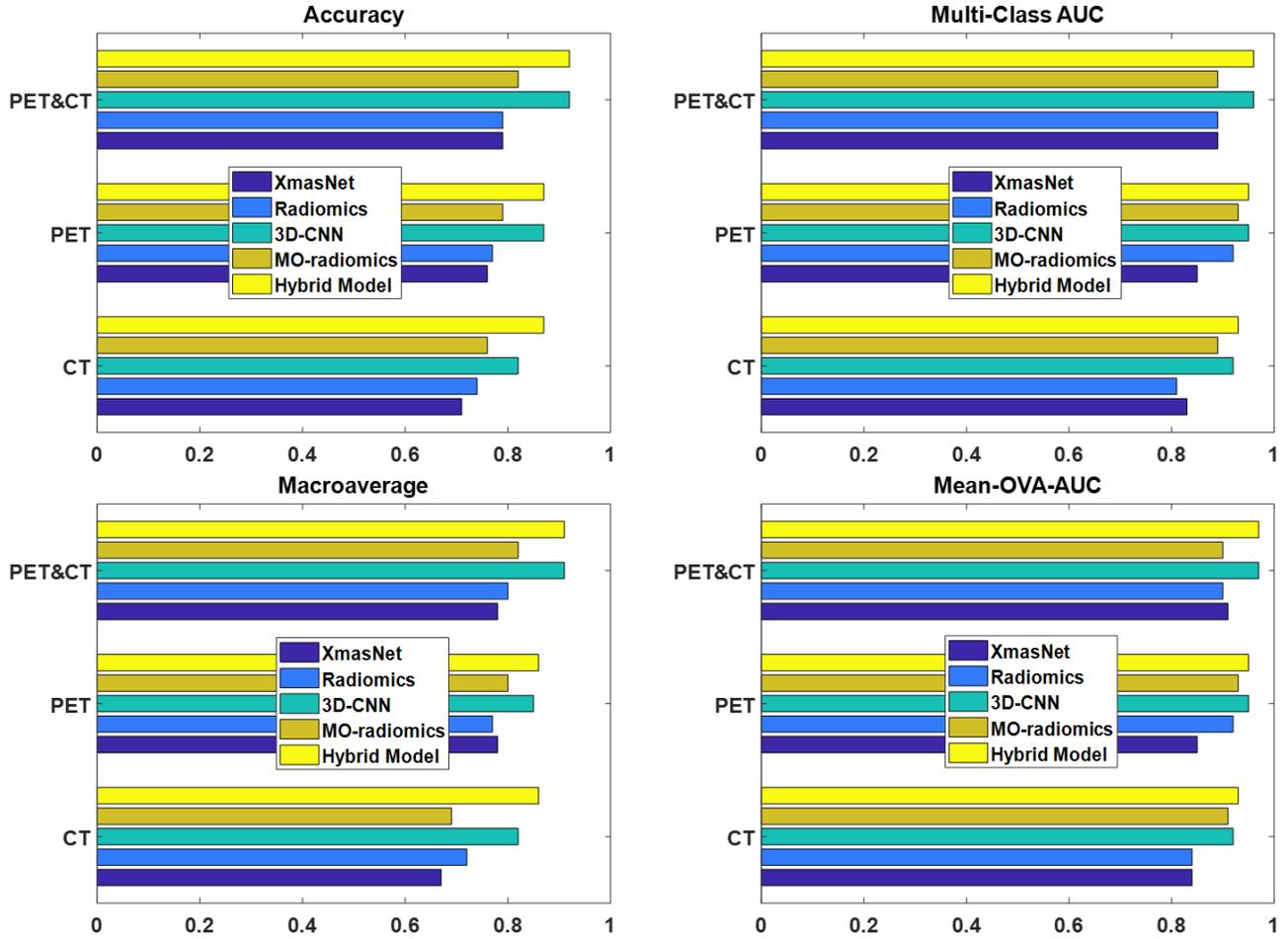

**Figure 3**: Bar plots of four evaluation criteria obtained by five methods. Longer bars indicate higher evaluation criteria values and better prediction results. The hybrid method that uses a combination of PET and CT images as input generated the best prediction results.

Confusion matrix results for the MO-radiomics model, the proposed 3D CNN model, and the hybrid model for the three categories of nodules are shown in Tables 5-7. All three confusion matrices show that differentiating suspicious nodes among three types of nodes is more difficult than differentiating normal or involved nodes. The hybrid model yielded better PA and UA than the MO-radiomics and CNN model when using CT imaging. The MO-radiomics model was worse in predicting suspicious nodes while better in predicting normal nodes than the CNN model by using CT imaging. After fusing the outputs of these two models by ER, the hybrid model improved both abilities of predicting normal and suspicious nodes. Moreover, when PET and CT imaging were combined, we obtained better PA and UA in most cases. Overall, our proposed predictive model was more effective than the two single models in predicting LNM. The same conclusion can be drawn from Figure 4. For the XmasNet and Radiomics models, it is difficult to find clear boundaries between each distribution of the prediction results for each type of node. However, the prediction results for the involved nodes obtained by the proposed CNN and MO-radiomics models can form a separately clustered region, which indicates that these two models better differentiate the involved nodes among three types of nodes than the XmasNet and Radiomics models. There is still no clear boundary between distributions of the prediction results for the normal and suspicious nodes for the proposed CNN and MO-radiomics models. The prediction results obtained by the hybrid method for three types of nodes construct three clustered regions, which implies that the hybrid method generates more reliable prediction results.

Table 5: Confusion matrix for MO-radiomics.

| Imaging | Node | Predicted Normal | Predicted Suspicious | Predicted Involved | UA |
|---|---|---|---|---|---|
| CT | *Normal* | 17 | 0 | 0 | 1 |
| | *Suspicious* | *3* | 2 | *4* | 0.22 |
| | *Involved* | *1* | *1* | 11 | 0.85 |
| | PA | 0.89 | 0.67 | 0.73 | |
| PET | *Normal* | 13 | *4* | 0 | 0.76 |
| | *Suspicious* | 0 | 7 | *2* | 0.78 |
| | *Involved* | *1* | *1* | 11 | 0.85 |
| | PA | 0.93 | 0.58 | 0.85 | |
| PET & CT | *Normal* | 14 | *3* | 0 | 0.82 |
| | *Suspicious* | 0 | 7 | *2* | 0.78 |
| | *Involved* | *1* | *1* | 11 | 0.85 |
| | PA | 0.93 | 0.64 | 0.85 | |

Table 6: Confusion matrix for 3D-CNN.

| Imaging | Node | Predicted Normal | Predicted Suspicious | Predicted Involved | UA |
|---|---|---|---|---|---|
| CT | *Normal* | 14 | *3* | 0 | 0.82 |
| | *Suspicious* | 0 | 7 | *2* | 0.78 |
| | *Involved* | 0 | *2* | 11 | 0.85 |
| | PA | 1 | 0.58 | 0.85 | |
| PET | *Normal* | 16 | *1* | 0 | 0.94 |
| | *Suspicious* | 0 | 7 | *2* | 0.78 |
| | *Involved* | *1* | *1* | 11 | 0.85 |
| | PA | 0.94 | 0.78 | 0.85 | |
| PET & CT | *Normal* | 17 | 0 | 0 | 1 |
| | *Suspicious* | 0 | 8 | *1* | 0.88 |
| | *Involved* | *2* | 0 | 11 | 0.85 |
| | PA | 0.89 | 1 | 0.92 | |

Table 7: Confusion matrix for the hybrid model.

| Imaging | Node | Predicted Normal | Predicted Suspicious | Predicted Involved | UA |
|---|---|---|---|---|---|
| CT | *Normal* | 16 | *1* | 0 | 0.94 |
| | *Suspicious* | 0 | 7 | *2* | 0.78 |
| | *Involved* | *1* | *1* | 11 | 0.85 |
| | PA | 0.94 | 0.78 | 0.85 | |
| PET | *Normal* | 16 | *1* | 0 | 0.94 |
| | *Suspicious* | 0 | 7 | *2* | 0.78 |
| | *Involved* | *1* | *1* | 11 | 0.85 |
| | PA | 0.94 | 0.78 | 0.85 | |
| PET & CT | *Normal* | 17 | 0 | 0 | 1 |
| | *Suspicious* | 0 | 8 | *1* | 0.88 |
| | *Involved* | *2* | 0 | 11 | 0.85 |
| | PA | 0.89 | 1 | 0.92 | |

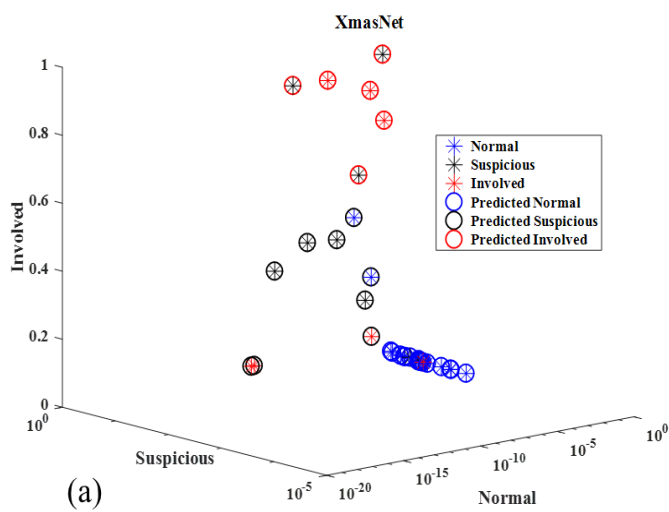

(a)

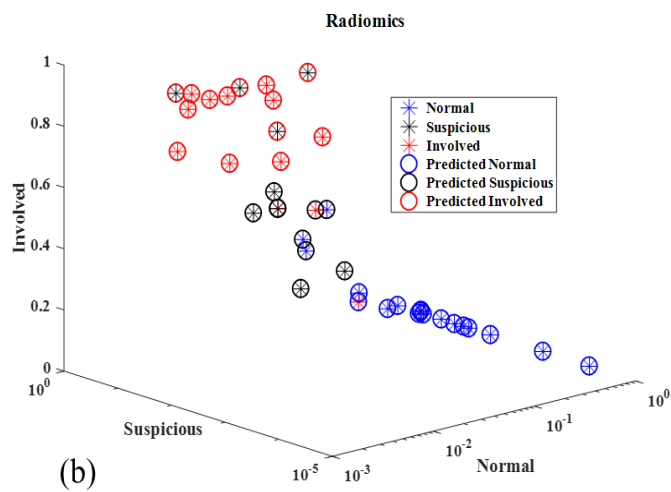

(b)

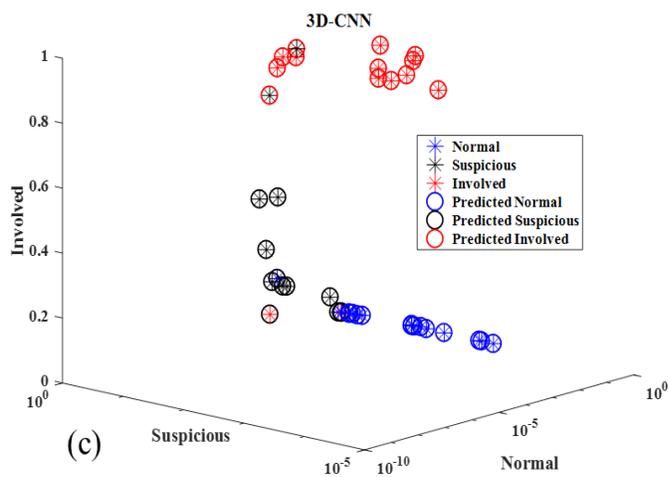

(c)

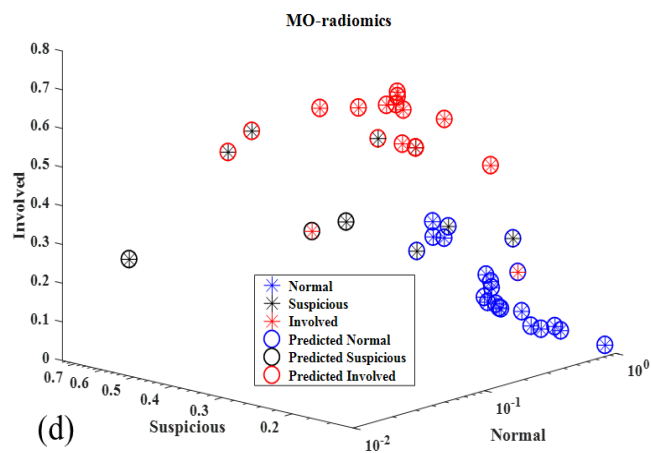

(d)

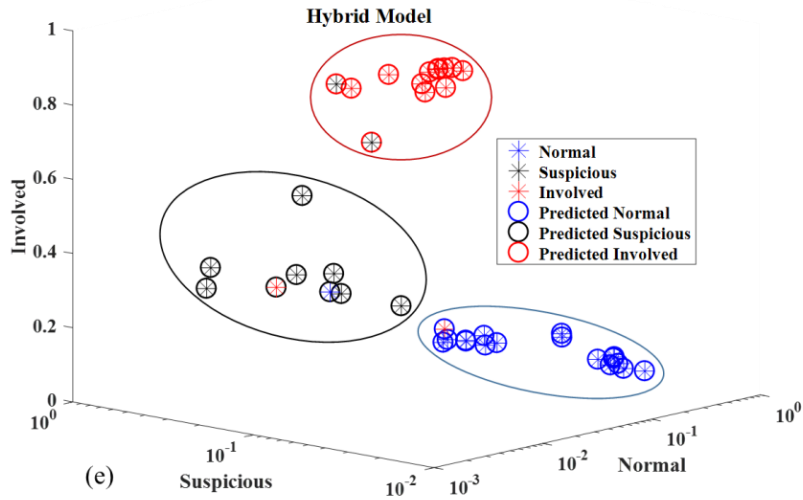

**Figure 4**: Prediction results obtained by five different models.

Finally, the four ROC curves in Fig. 5 illustrate the hybrid method's performance in distinguishing different types of nodes by using combination of CT and PET imaging. The hybrid method achieved 0.98 AUC when differentiating normal nodes from the other two types of nodes (suspicious and involved). Additionally, the hybrid model achieved 0.99, 0.97, and 0.82 AUC values for distinguishing normal from suspicious, normal from involved, and suspicious from involved nodes, respectively.

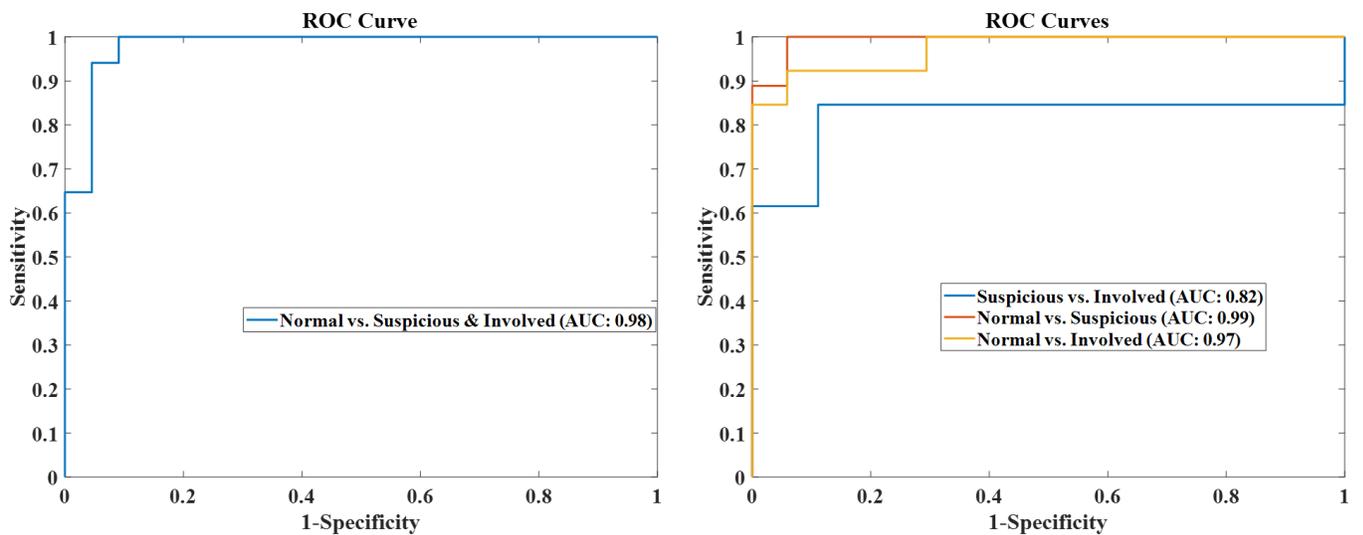

**Figure 5**: ROC curves for the Hybrid model.

## 4. Conclusion

We proposed a hybrid model that predicts lymph node metastasis in head and neck cancer by combining outputs of MO-radiomics and 3D-CNN models through an evidential reasoning fusion approach. Specifically, to obtain more reliable performance, we developed a new MO-radiomics model based on our previous work. This new model considers PAs and UAs in confusion matrix as objective functions, in addition to sensitivity and specificity. Meanwhile, we developed a 3D-CNN model to make full use of contextual information in the images. The final output was obtained by combining the two models' outputs using the ER approach. The experimental results show that the hybrid model improved the classification accuracy and reliability obtained by the two single models when using CT imaging alone. We also investigated the influence of input imaging. We obtained better results using both PET and CT imaging than using PET or CT imaging alone.

The current MO-radiomics model optimized PAs and UAs simultaneously. These two types of objective function can be trained alternately, which could potentially improve the model's performance. To obtain a more robust model, transfer learning can be introduced into the 3D-CNN model as a next step. The dataset can also be expanded to include more patient data to build and validate the model, so it can be applied in the clinical settings. With better prediction of type of lymph nodes, we can make a better individualized treatment plan, potentially resulting in better control and lower toxicity in the HNC radiation treatment.

## 5. Acknowledgment

This work was supported in part by the American Cancer Society (ACS-IRG-02-196) and the US National Institutes of Health (R01 EB020366). The authors would like to thank the following people for their contributions to this manuscript: Dr. Genggeng Qin from Southern Medical University, Guangzhou, China and Dr. Jonathan Feinberg from UT southwestern Medical Center.